\pdfoutput=1
\documentclass[twocolumn,
floatfix,preprintnumbers,citeautoscript,prl,superscriptaddress,amsmath,amssymb]{revtex4}

\usepackage[utf8]{inputenc}
\usepackage{graphicx}
\usepackage{hyperref}
\usepackage{color}

\begin{document}

\title{Is Gravity the Weakest Force?}

\preprint{IPMU19-0062}

\date{April, 2019}

\author{Satoshi Shirai}
\affiliation{Kavli Institute for the Physics and Mathematics of the Universe (WPI), University of Tokyo Institutes for Advanced Study, University of Tokyo, Chiba 277-8583, Japan}

\author{Masahito Yamazaki}
\affiliation{Kavli Institute for the Physics and Mathematics of the Universe (WPI), University of Tokyo Institutes for Advanced Study,  University of Tokyo, Chiba 277-8583, Japan}

\date{April, 2019}

\begin{abstract} 
It has recently been suggested that ``gravity is the weakest force'' in any theory with a suitable UV completion within quantum gravity. One formulation of this statement is the scalar weak gravity conjecture, which states that gravity is weaker than the force originating from scalar fields. We study the scalar weak gravity conjecture in de Sitter space, and discuss its low-energy consequences in light of the experimental searches for fifth forces and violations of the equivalence principle. We point out that some versions of the scalar weak gravity conjecture forbid the existence of very light scalar particles, such as the quintessence and axion-like particles. The absence of the quintessence field means that these versions of the scalar weak gravity conjecture are in phenomenological tension with the recently-proposed de Sitter swampland conjecture and its refinements. Some other versions of the scalar weak gravity conjecture escape these constraints, and could have interesting phenomenological consequences.
\end{abstract}

\maketitle

\bigskip\noindent
{\bf Introduction}

Gravity has been studied extensively in physics for centuries.
Yet despite its long history of research, gravity is still
one of the most enigmatic phenomena of Nature.
The unification of gravity with all other forces of Nature,
namely the quantum gravity, has been a notoriously difficult subject,
and gravity seems to be intrinsically different from anything else.

One of the rather peculiar features of gravity is that it is extremely weak. We learn this in kinder garden when
we find that the electromagnetic force of a small magnet wins over the gravity of the whole earth.
Recently this simple observation, that  ``gravity is the weakest force'', has been promoted to a principle
in the context of the swampland program \cite{Vafa:2005ui,Ooguri:2006in}, a program to work out the low-energy consequences of the existence of UV completions with gravity.

The most famous formulation of the weakness of the gravity is the 
weak gravity conjecture (WGC) \cite{ArkaniHamed:2006dz}, which
states the following \footnote{This is the so-called electric version of the WGC. We will not discuss the magnetic version of the WGC in this Letter.}:
in a theory with a consistent UV completion in theories of quantum gravity,
there should exist a particle
with mass $m$ and $U(1)$ gauge charge $q$ satisfying the inequality:
\begin{align}
     \sqrt{2} \, q \, e \ge \frac{m}{M_{\rm Pl}} ,
     \label{GWGC}
\end{align}
where $M_{\rm Pl}\simeq 2.4\times 10^{18} \, \textrm{GeV}$ is the reduced Planck mass
and $e$ is the gauge coupling constant for the $U(1)$ gauge symmetry.
While this is still a conjecture, there has been good arguments supporting this conjecture
from the decay of non-extremal black holes \cite{ArkaniHamed:2006dz}, and 
there has recently been many more supporting arguments (see e.g.\ Refs.~\cite{Cheung:2014ega,Harlow:2015lma,Cottrell:2016bty,Hod:2017uqc,Fisher:2017dbc,Crisford:2017gsb,Cheung:2018cwt,Hamada:2018dde,Urbano:2018kax}). \footnote{Once should notice, however, there are several different versions of the WGC 
in the literature. For example, in one version the inequality \eqref{GWGC} is satisfied by a black hole, rather than a particle.}

The inequality \eqref{GWGC} means that the force mediated by the gauge boson (spin $1$ particle) is stronger by that by the graviton (spin $2$ particle), namely $F_{\rm gauge}\ge  F_{\rm gravity}$ with 
\begin{align}
&F_{\rm gauge}=\frac{(q e)^2}{4\pi r^2}  ,  \quad F_{\rm gravity}=\frac{m^2}{8\pi M_{\rm Pl}^2 r^2}.
\end{align}
 This raises a natural question: what happens if we have a spin $0$ particle, namely a scalar? Is gravity still the weakest force, and if so, how should we articulate this condition?

There has recently been several attempts towards answering this question, by formulating a set of 
``scalar versions'' of the WGC \cite{Palti:2017elp,Lust:2017wrl,Lee:2018urn,Palti:2019pca,Gonzalo:2019gjp}. However, the contrast with the case of the original weak gravity conjecture (which we call the gauge WGC or GWGC, to be distinguished from the scalar WGC or SWGC),
SWGCs have much less evidence in general, and it is often not clear
exactly which versions of the conjecture should be adopted.

For this reason we work out low-energy consequences of several versions of the WGC,
and discuss observational constraints on them. Such bottom-up constraints on 
swampland conjectures have been extremely useful in sharpening our understanding of 
quantum gravity (see Ref.~\cite{Yamazaki:2019ahj} for recent summary), and this Letter is not an exception---we will draw interesting conclusions, including some phenomenological tension with the de Sitter swampland conjectures.

\bigskip\noindent
{\bf Scalar Weak Gravity Conjecture}

Let us begin by stating one version of the SWGC \cite{Palti:2017elp} (we will later comment on 
 variations of the conjecture).
Suppose a charged particle has a mass of $m$.
This particle is arbitrary and not necessarily an elementary particle as long as it is a GWGC state, namely if it satisfies the inequality \eqref{GWGC}. 

Suppose that the mass $m$ depends on a set of exactly massless scalar fields $\varphi^i$.
This means that the GWGC states
have trilinear couplings with the massless fields $\varphi^i$.
To see this, suppose that the GWGC state is a complex scalar $\phi$. We then have a 
field-dependent mass term $m^2(\varphi) |\phi|^2$ in the Lagrangian $\mathcal{L}$, and when expanded around a VEV (vacuum expectation value)
$\varphi_0$ of the field $\varphi$ as $\varphi=\varphi_0+\delta \varphi$, 
we obtain a trilinear coupling 
\begin{align}
\mathcal{L}\supset
\partial_{\varphi} m^2(\varphi_0) (\delta \varphi) |\phi|^2.
\label{trilinear}
\end{align}
The case of a fermion is similar. 

Let us denote the metric for the kinetic term of $\varphi^i$ to be $g_{ij}$ (which can depend on $\varphi$ themselves), with the Lagrangian ${\cal L}_{\rm kin} = - g_{ij} \partial \varphi^{i}\partial \varphi^{j}$.
Then the SWGC states that we have an inequality
\begin{align}
 |\partial_{\varphi} m|^2 \equiv
  \sum_{i,j} g^{ij} (\partial_{\varphi_i} m)  (\partial_{\varphi_j} m) \ge \frac{m^2}{M_{\rm Pl}^2}  \;.
    \label{SWGC}
\end{align}
The physical content of this inequality is that the total force mediated by the massless fields $\varphi_i$ is stronger than that by gravity, when we consider the $2\to 2$ scattering of the GWGC states. Namely, we have $F_{\rm scalar}\ge  F_{\rm gravity}$ with
\begin{align}
&F_{\rm scalar}=\frac{ |\partial_{\varphi} m|^2 }{4\pi r^2}  , \quad
F_{\rm gravity}=\frac{m^2}{8\pi M_{\rm Pl}^2 r^2}.
\end{align}
Note that the trilinear coupling as in \eqref{trilinear}
generates the scalar force as on the left hand of \eqref{SWGC}.

\bigskip\noindent
{\bf Scalar Weak Gravity Conjecture in de Sitter Space}

In this Letter, we wish to apply the SWGC to our Universe,
namely to a de Sitter space with a positive value of the cosmological constant
and a positive value of the Hubble constant $H$.
While swampland conjectures are often formulated as theoretical constraints on possible low-energy physics
and does not refer to the history of the Universe, our considerations in this Letter are more physical,
and our intention is to honestly formulate the statement that ``gravity is the weakest force'' in our Universe.

In this context, we claim that we should allow the fields $\varphi_i$ to be not exactly massless, as long as they are nearly massless. More quantitatively, we allow the mass comparable to the 
$m_{\varphi_i}$ are of order of the value of the Hubble constant of the Universe,
$H$: $m_{\varphi_i} \lesssim H$,
and on the left hand side of \eqref{SWGC} we sum over 
all such fields.

There are several motivations for allowing such nearly-massless fields in the sum.
First, when we consider $2\to 2$ scattering of such particles inside the Universe
the nearly-massless fields $\varphi_i$ are practically massless, since their Compton wavelength is 
comparable or larger than the current horizon scale $\sim H^{-1}$, where the scattering takes place.
Second, in the Universe with a positive value of the cosmological constant
and one then expects that the field $\phi$ 
will generically obtain a mass of the order $H$ from curvature couplings.

Unlike the supersymmetric case, it is generally hard to keep a scalar particle massless in the de Sitter space.
Even when the fields $\varphi_i$ are massless classically, 
we expect that the one-loop effect from the trilinear coupling \eqref{trilinear} will generate a mass for $\varphi$,
unless there is some symmetry reasons. Note that we cannot forbid such a mass term by imposing 
an exact global symmetry, since global symmetry is not allowed in 
theories of quantum gravity (unless it is emergent in the IR) \cite{Misner:1957mt,Polchinski:2003bq,Banks:2010zn}. \footnote{One of the motivations for the SWGC comes from black holes in four-dimensional $\mathcal{N}=2$ supersymmetry \cite{Palti:2017elp},
where such massless scalar fields are present. It is difficult in general, however, to 
ensure the presence of such massless scalar fields in non-supersymmetric theory.}

When applied to the present-day Universe, the masses of the $\varphi_i$
are extremely small:
\begin{align}
    m_{\varphi_i} \lesssim H_0 \sim O(10^{-33})~ \textrm{eV} .
    \label{mass_ineq}
\end{align}

As this discussion makes clear, we impose weakness of gravity in 
at the IR, namely the horizon scale. While there could be other fields with masses much larger than 
the nearly-massless fields $\varphi_i$, they create only short-range forces
at the horizon scale and hence do not affect our argument, as in the case of the GWGC \eqref{GWGC}.

\bigskip\noindent
{\bf Constraints on Very Light Scalars}

In the Standard Model of particle physics, there are no nearly-massless scalar fields 
whose masses satisfy \eqref{mass_ineq}. Such particles, however,
often arise in extensions of the Standard Model.

One scenario for a nearly-massless scalar is an ultralight axion-like particle (ALP).
It has been argued that string theory motivates the presence of multiplet ALPs across all energy scales \cite{Arvanitaki:2009fg}, possibly including those satisfying \eqref{mass_ineq}.
Another example is the quintessence field \cite{Ratra:1987rm,Wetterich:1987fm,Zlatev:1998tr}, a dynamical scalar field for the dark energy. Such a scalar field should be nearly massless and satisfy \eqref{mass_ineq}, to avoid rapid change of the size of the dark energy.
More generally, string theory has many moduli, and some of these fields could have flat directions
which are only broken by non-perturbative effects. Of course, these possibilities are not mutually exclusive,
since part of the moduli could generate an ALP, which could play the role of the quintessence, for example.

Let us consider a minimal extension of the Standard Model
where we only one very light scalar particle satisfying 
\eqref{mass_ineq}, and let us denote this scalar by $\varphi$.
We choose a canonical kinetic term for this scalar field, by a suitable redefinition of the field if needed.

Let us apply the SWGC to our setup. First, we have the choice of GWGC states, but 
there is practically little constraint from this condition. As stated in introduction,
gravity is extremely weak compared to other forces and hence all charged states in the Standard Model are GWGC states.

We obtain a constraint on a single function $m(\varphi)$ from the SWGC:
\begin{align}
2 \left| \partial_{\varphi} m  \right|^2 \ge \left( \frac{m}{M_{\rm Pl}}\right)^2 .
\label{WGC_applied}
\end{align}

This inequality \eqref{WGC_applied} represents the presence of the fifth force,
which acts stronger than gravity in any charged states, e.g., the electron and the proton.

Such a scenario is highly constrained in observational constraints on fifth force searches (see e.g., Refs.~\cite{Fischbach:1999bc,Bertotti:2003rm,EspositoFarese:2004cc,Will:2005va}).
First, one generically expects that the 
coupling of the field $\varphi$ (as in eq.~\eqref{trilinear}) to other GWGC states are non-universal.
This causes the violation of the weak equivalence principle, which has been checked to be 
very high accuracy (of order $10^{-13}$). 
The constraint is ameliorated when we somehow manage to couple the light field $\varphi$
universally to the matters as the gravity. It is still the case, however, that $\varphi$ 
modifies the gravitational potential, leading to sizable deviations of the 
gravitational potential in the parametrized post-Newtonian expansion.
The observational constraint from this reads
\begin{align}
\left| \partial_{\phi} m  \right|^2 < O(10^{-5})\left( \frac{m}{M_{\rm Pl}}  \right)^2.
\label{fifth_force_constraint}
\end{align}
This is in immediate tension with the SWGC \eqref{WGC_applied}. \footnote{It would be interesting to see
if we can evade these constraints by using some screening mechanisms, e.g.\ Vainshtein \cite{Vainshtein:1972sx} and Chameleon \cite{Khoury:2003rn} mechanisms. Such mechanisms can have interesting implications for future observations. Of course, even if we succeed in this, it is a separate and highly non-trivial question if such options can be reproduced from theories of quantum gravity (see e.g.\ \cite{Hinterbichler:2010wu} for such an attempt). Note that some of the screening scenarios are accompanied by superluminal propagations, which are obstructions for UV completion \cite{Adams:2006sv}.}

The intuitive reason for our findings is clear---SWGC states that the fifth force originating from the scalar force should be stronger than gravity, which is already excluded from fifth-force searches.

Let us add that if $\phi$ has a potential and the VEV of $\phi$ has a time dependence, the masses of particles are time-dependent, and such time-variations are also strongly constrained by 
 cosmological observations.

Summarizing, we have concluded that the SWGC as formulated above is in tension with the existence of 
a very light scalar. The case of multiple such scalars is similar.

\bigskip\noindent
{\bf Tension with de Sitter Swampland Conjecture}

The results we have presented above has an interesting consequence: the SWGC 
is in phenomenological tension with another swampland conjecture, the de Sitter swampland conjecture  \cite{Obied:2018sgi}.

Let us quickly summarize the de Sitter swampland conjecture.
After the initial proposal \cite{Obied:2018sgi},
some bottom-up constraints of the conjecture has been pointed out in Refs.~\cite{Denef:2018etk,Conlon:2018eyr,Murayama:2018lie,Choi:2018rze,Hamaguchi:2018vtv}, leading to several proposals for 
refinements \cite{Dvali:2018fqu,Andriot:2018wzk,Garg:2018reu,Murayama:2018lie,Ooguri:2018wrx,Garg:2018zdg,Andriot:2018mav}. In particular, 
Refs.~\cite{Garg:2018reu,Ooguri:2018wrx} proposed that a scalar potential $V$ 
for a low-energy effective field theory
should satisfy the inequality 
\begin{align}
M_{\rm Pl}\, |\nabla V|>c \, V   \textrm{ or } \,
M_{\rm Pl}^2\, \textrm{min}(\nabla\nabla  V) \leq - c' V.
\label{conjecture}
\end{align}
Here $c$ and $c'$ are $O(1)$ positive constants (a version with $c'=0$ was 
proposed in Ref.~\cite{Murayama:2018lie}).

Irrespective of the details of the refinements, all of these conjectures imply that the (meta)stable de Sitter space
($\nabla V=0, V>0$) are excluded  (see also Refs.~\cite{Dine:1985he,Maldacena:2000mw,Steinhardt:2008nk,McOrist:2012yc,Sethi:2017phn,Danielsson:2018ztv} for related discussion). This gives a rather strong motivation for quintessence models as an explanation of the dark energy. This, as we have seen, contradicts the SWGC.

This in itself does not prove the inconsistency between SWGC and the (refined) de Sitter swampland conjectures, since one can try to explain dark energy by modify the gravity instead of incorporating 
extra quintessence matter fields. However, many modifications of the gravity, such as  
scalar-tensor theories, have light scalar fields and tend to have issues similar to the quintessence fields.
Moreover, it is a non-trivial question to see if such modifications of gravity can really be embedded into
theories of quantum gravity, such as string theory.

We therefore come to the conclusion that the version of the SWGC discussed above is in tension
with the (refined) de Sitter swampland conjecture. 
This is a non-trivial result. Indeed, both conjectures
are partly motivated by the same conjecture, namely the swampland distance conjecture \cite{Ooguri:2006in,Klaewer:2016kiy}, at least in asymptotic regions of the parameter space. 
Moreover it was suggested in Ref.~\cite{Ooguri:2018wrx}
that there are close analogies between the two conjectures, when the scalar potential is given by the 
mass itself. More broadly, consistency checks between 
different conjectures has been one of the 
primary guidelines in the swampland program.
It is fair to say that neither conjectures have solid evidence, and are
open for possible modifications and refinements.
Since we already mentioned on refinements on the de Sitter swampland conjectures,
let us next discuss variations of the SWGC.

\bigskip\noindent
{\bf Variations of the Scalar Weak Gravity Conjecture}

Let us discuss variations of the SWGC---when stating the inequality
\eqref{mass_ineq} there are many choices. Some variations, which impose stronger 
condition than above, clearly does not affect
the argument  above. For example, 
we can impose the inequality for arbitrary states,
as opposed to only for GWGC particles.
We  can also replace the inequality ($\ge$) by a strict inequality ($>$) or an approximate inequality ($\gtrsim$),
without affecting our argument.

One plausible modification of the SWGC is to consider the following:
instead of imposing the inequality \eqref{SWGC} for any state (or any GWGC state),
we can require that {\it there exists at least one particle } (the SWGC particle)  satisfying the SWGC condition  \eqref{SWGC}.
This extends the ``weak version'' of the SWGC in Ref.~\cite{Palti:2017elp} to the de Sitter space,
and represents most directly the spirit that the gravity is the weaker
than the other forces. \footnote{Yet another version of the conjecture is to 
claim the existence of a particle for which we have the inequality 
$|F_{\rm gauge}|\ge |F_{\rm scalar}|+|F_{\rm gravity}|$ \cite{Palti:2017elp}. We do not discuss this version in this Letter. Morally speaking this version should be stated as ``gauge as the strongest force'', rather than ``gravity as the weakest force''.}

Note we still assume that there exists at least one particle satisfying the GWGC condition  \eqref{GWGC},
but this particle can in general be different from the particle satisfying the SWGC. 
We can impose a stronger constraint that both inequalities are satisfied by the same particle.

In these versions of the SWGC, if the mass $m$ of the SWGC particle satisfying the inequality \eqref{SWGC} is extremely large (e.g.\ near the cutoff scale of theory),
then it would likely be difficult to constrain the existence of such particles observationally.
One possible exception is the case
where such a particle is stable and is a dark matter candidate.

In addition to the SWGC particle, we discussed very light scalar fields $\varphi_i$.
One should note that morally speaking  the conjecture asks for the existence of a very light scalar particle. While we {\it assumed} the existence of such a particle in the discussion above in the formulation of the SWGC,  the idea that gravity is the weakest force at least at the horizon scale seems to {\it require}
such a particle--without such a light scalar particle the scalar force decays in the IR.
While this light scalar field does not need to couple directly to the Standard Model particles,
it will necessarily couple to them through gravitational interactions. 

\bigskip\noindent
{\bf Strong Scalar Weak Gravity Conjecture}

We have to this point formulated SWGC in the IR. 
One might however try to consider a 
``mixed UV/IR version'' of the SWGC. In this situation, 
many states, including massive states, begin to contribute to the forces, and a priori 
we need to take into account all of them, and one might wonder if there is any hope for 
a simple statement, while keeping the statement applicable to any low-energy effective theory
with UV completions.

Despite these potential obstacles, recently the authors of Ref.~\cite{Gonzalo:2019gjp}
boldly conjectured a version of the SWGC, which we call the strong scalar WGC (SSWGC) (see also \cite{Brahma:2019mdd}). \footnote{This should not be confused with some other versions of the SWGC discussed before, which are sometimes referred to as ``strong'' versions of the SWGC.}

The SSWGC states that the scalar potential $V$, in a low-energy theory with UV completion with gravity, satisfies an inequality
\begin{align}
\chi \equiv 2 (V''')^2- V'' V''''\ge  \frac{(V'')^2}{M_{\rm Pl}^2}.
\label{SSWGC}
\end{align}

If we write $m^2=V''$, then this inequality can be written as 
\begin{align}
2  (\partial_{\phi} m^2)^2- m^2 (\partial^2_{\phi} m^2) \ge \frac{(m^2)^2}{M_{\rm Pl}^2} .
\label{SSWGC_m}
\end{align}
which looks similar to \eqref{SWGC}.

There are, however, crucial differences between  \eqref{SWGC}
and \eqref{SSWGC}. First, as already stated the conjecture \eqref{SSWGC_m} is 
meant to be a ``mixed UV/IR statement'' \cite{Gonzalo:2019gjp}, 
while we have above been using the versions formulated in  the IR.
Second, \eqref{SSWGC_m} has a term involving the fourth derivative of the potential,
which did not appear in \eqref{SWGC}. 
Third, the new proposal \eqref{SSWGC} does not 
refer to existence of nearly-massless or massless fields ($\varphi_i$ in our previous notation),
and refers only to the scalar field $\phi$, which can be taken arbitrary. 

In SSWGC the self-interaction of the scalar field, without any help from other scalars,
is claimed to win over gravity. We find a tower of massless states when the equality in \eqref{SSWGC},
which states are proposed to play the role of the nearly-massless states ($\varphi_i$ in the notation of this Letter). In this situation,
since we have a tower of very light scalar fields in the theory, 
we expect that our conclusions presented above from the fifth-force constraints seems to apply essentially the same manner to this new conjecture. However, such a tower often arises 
when the field range is of order the Planck scale, where the effective field theory breaks down
according to the distance conjecture \cite{Ooguri:2006in,Klaewer:2016kiy}.
For this reason SSWGC could possibly escape our observational constraints,
despite the fact that the conjecture applies to any scalar.

One  obvious question is of course if the SSWGC is true, at least in some corners of the landscape of 
quantum gravity. While we leave this question for future work, let us consider a simplest setup 
where we have a compactification of the eleven-dimensional supergravity on a Calabi-Yau four-fold.
This is considered to be the classical limit of M-theory compactifications, and we assume that 
all the corrections, stringy and $\alpha'$, can be safely neglected.

The bosonic part of the action of the eleven-dimensional supergravity reads
\begin{align}
S=\frac{1}{2\kappa_{11}^2} \int d^{11}  x\sqrt{-g^{11}} \left(
R- \frac{1}{2} |G|^2
\right),
\label{S11}
\end{align}
where $R$ is the curvature and $G$ is the four-form field strength.
Let us compactify this theory on $\mathbb{R}^4\times (\textrm{7-manifold})$,
and denote the overall modulus of the compactification manifold by $\rho$.
As explained in \cite{Obied:2018sgi}, the effective potential
for the overall modulus $\rho$ takes the form
\begin{align}
V_{\rm eff}= V_{R} \, e^{- \frac{6}{\sqrt{14}} \frac{\rho}{M_{\rm Pl}}}
+V_{G} \, e^{- \frac{10}{\sqrt{14}} \frac{\rho}{M_{\rm Pl}}} .
\label{Veff}
\end{align}
The two terms represents the contributions from the curvature and the 
four-form field strength in \eqref{S11}, and we hence have $V_G\ge 0$.
Interestingly, this seems to satisfy \eqref{SSWGC}, irrespective of the sign of $V_R$.

Another question, which is natural for this Letter, is if the SSWGC holds for the Standard Model of particle physics, e.g.\ the Higgs field. \footnote{This question was previously discussed in \cite{Gonzalo:2019gjp}, although their analysis is limited to the classical Higgs potential.}

\begin{figure}[htp]
\includegraphics[width=0.45\textwidth]{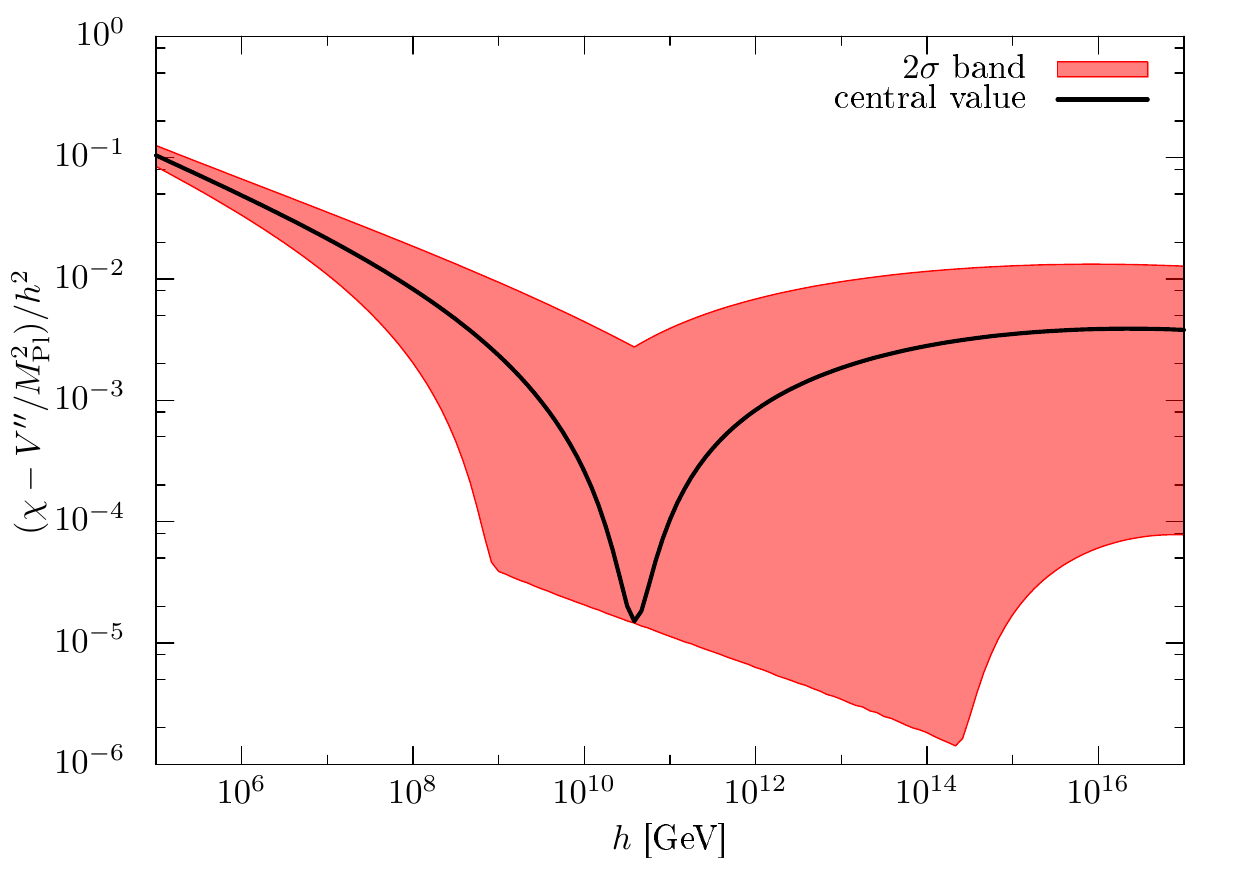}
\caption{The plot of $(\chi - (V'')^2/M_{\rm Pl}^2)/h^2$ for the Standard Model Higgs effective potential.
The SSWGC states that this quantity is non-negative.}
\label{fig:higgs}
\end{figure}

For the Standard Model Hversioniggs, the effective potential of the Higgs boson $h$, is approximately given by
\begin{align}
    V(h) = \frac{\lambda_{\rm eff}(h)}{4} h^4, \label{higgs}
\end{align}
for $h$ much greater than the electroweak scale.
The SSWGC criteria \eqref{SSWGC} for the Higgs boson is approximated as,
\begin{align}
    \chi(h) \simeq \frac{3h^2}{8}\left[ 144 (\lambda_{\rm eff}(h) + \beta_{\rm eff}(h))^2 + 23 \beta_{\rm eff}^2(h)   \right],
\end{align}
where $\beta_{\rm eff}(h) \equiv d \lambda_{\rm eff}/ d \log(h)$ and we neglect higher derivatives on $\beta_{\rm eff}$, as these values are higher-loop suppressed.
Unless both $\lambda$ and $\beta_{\rm eff}$ are zero, the value of $\chi$ is positive and greater than  $O(h^2/(16\pi^2)^2)$.
Therefore, in the Standard Model, the value of $\chi$ is greater than the Planck-suppressed combination $V''/M_{\rm Pl}^2$.
In Fig.~\ref{fig:higgs}, we show the $\chi/h^2 - V''/M_{\rm Pl}^2$ as a function of the Higgs VEV $h$.
The red band in the figure shows $2\sigma$ uncertainty from the experimental and theory errors.
To calculate the effective potential, we use the procedure of Refs.~\cite{Degrassi:2012ry,Buttazzo:2013uya} and take the physical parameters as top mass $m_{t} = 173.0\pm0.4$ GeV and Higgs mass $m_h=125.18 \pm 0.16 $ GeV \cite{Tanabashi:2018oca}.
We add another error $\pm 0.5$ GeV to the top quark mass due to the hadronic uncertainty.
For the other parameters, we use same values as in Ref.~\cite{Nagata:2013sba}.

Note that we use the inequality \eqref{SSWGC} from version 2 of Ref.\,\cite{Gonzalo:2019gjp}.
In version 1 of Ref.\,\cite{Gonzalo:2019gjp}, the criteria is $\chi/V''\geq 1/M_{\rm Pl}^2$, so that 
the direction of this inequality is opposite from that in version 2 when $V''<0$. 
The criteria in version 1 is violated by the potential \eqref{Veff}, if $\rho > M_{\rm Pl} \log(-25 V_G/9 V_R)/\sqrt{7/8}$ with $V_R<0$ and the Higgs potential \eqref{higgs} with $h \gtrsim 10^{10}$ GeV.

As discussed before, the some versions of the SWGC are in tension with the 
existence of very light scalars. Since the Higgs or meson scalar field mediates forces much larger than the gravity, this tension may be relaxed in the SSWGC condition \eqref{SSWGC}, if we suitably generalize the conjecture to multi-field cases. Of course, the fundamental question remains how to better motivate the SSWGC in itself.

In conclusion,  SWGC is an attempt to articulate precise our intuition that
gravity is the weakest force in Nature. Theoretically, however, it is not clear 
which version of the conjecture should hold. We have seen that observational constraints on fifth-force searches 
provide useful guidelines in formulating the SWGC. Our considerations highlights the 
fascinating link between theoretical studies of ``high-scale'' quantum gravity and 
experimental studies of yet unknown ``low-scale'' physics.

\bigskip\noindent
{\bf Acknowledgements}

We would like to thank Ryo Saito for useful discussions and for comments on the manuscript. MY would like to thank Perimeter institute for hospitality where part of this work was performed. This research was supported in part by WPI Research Center Initiative, MEXT, Japan (SS, MY) and by the JSPS Grant-in-Aid for Scientific Research 17H02878 (SS), 18K13535 (SS), 19H04609 (SS), 17KK0087 (MY), 19K03820 (MY), and 19H00689 (MY).

\bibliographystyle{apsrev4-1}
\bibliography{swampland_scalar_bib}

\begin{thebibliography}{57}%
\makeatletter
\providecommand \@ifxundefined [1]{%
 \@ifx{#1\undefined}
}%
\providecommand \@ifnum [1]{%
 \ifnum #1\expandafter \@firstoftwo
 \else \expandafter \@secondoftwo
 \fi
}%
\providecommand \@ifx [1]{%
 \ifx #1\expandafter \@firstoftwo
 \else \expandafter \@secondoftwo
 \fi
}%
\providecommand \natexlab [1]{#1}%
\providecommand \enquote  [1]{``#1''}%
\providecommand \bibnamefont  [1]{#1}%
\providecommand \bibfnamefont [1]{#1}%
\providecommand \citenamefont [1]{#1}%
\providecommand \href@noop [0]{\@secondoftwo}%
\providecommand \href [0]{\begingroup \@sanitize@url \@href}%
\providecommand \@href[1]{\@@startlink{#1}\@@href}%
\providecommand \@@href[1]{\endgroup#1\@@endlink}%
\providecommand \@sanitize@url [0]{\catcode `\\12\catcode `\$12\catcode
  `\&12\catcode `\#12\catcode `\^12\catcode `\_12\catcode `\%12\relax}%
\providecommand \@@startlink[1]{}%
\providecommand \@@endlink[0]{}%
\providecommand \url  [0]{\begingroup\@sanitize@url \@url }%
\providecommand \@url [1]{\endgroup\@href {#1}{\urlprefix }}%
\providecommand \urlprefix  [0]{URL }%
\providecommand \Eprint [0]{\href }%
\providecommand \doibase [0]{http://dx.doi.org/}%
\providecommand \selectlanguage [0]{\@gobble}%
\providecommand \bibinfo  [0]{\@secondoftwo}%
\providecommand \bibfield  [0]{\@secondoftwo}%
\providecommand \translation [1]{[#1]}%
\providecommand \BibitemOpen [0]{}%
\providecommand \bibitemStop [0]{}%
\providecommand \bibitemNoStop [0]{.\EOS\space}%
\providecommand \EOS [0]{\spacefactor3000\relax}%
\providecommand \BibitemShut  [1]{\csname bibitem#1\endcsname}%
\let\auto@bib@innerbib\@empty
\bibitem [{\citenamefont {Vafa}(2005)}]{Vafa:2005ui}%
  \BibitemOpen
  \bibfield  {author} {\bibinfo {author} {\bibfnamefont {C.}~\bibnamefont
  {Vafa}},\ }\href@noop {} {\  (\bibinfo {year} {2005})},\ \Eprint
  {http://arxiv.org/abs/hep-th/0509212} {arXiv:hep-th/0509212 [hep-th]}
  \BibitemShut {NoStop}%
\bibitem [{\citenamefont {Ooguri}\ and\ \citenamefont
  {Vafa}(2007)}]{Ooguri:2006in}%
  \BibitemOpen
  \bibfield  {author} {\bibinfo {author} {\bibfnamefont {H.}~\bibnamefont
  {Ooguri}}\ and\ \bibinfo {author} {\bibfnamefont {C.}~\bibnamefont {Vafa}},\
  }\href {\doibase 10.1016/j.nuclphysb.2006.10.033} {\bibfield  {journal}
  {\bibinfo  {journal} {Nucl. Phys.}\ }\textbf {\bibinfo {volume} {B766}},\
  \bibinfo {pages} {21} (\bibinfo {year} {2007})},\ \Eprint
  {http://arxiv.org/abs/hep-th/0605264} {arXiv:hep-th/0605264 [hep-th]}
  \BibitemShut {NoStop}%
\bibitem [{\citenamefont {Arkani-Hamed}\ \emph {et~al.}(2007)\citenamefont
  {Arkani-Hamed}, \citenamefont {Motl}, \citenamefont {Nicolis},\ and\
  \citenamefont {Vafa}}]{ArkaniHamed:2006dz}%
  \BibitemOpen
  \bibfield  {author} {\bibinfo {author} {\bibfnamefont {N.}~\bibnamefont
  {Arkani-Hamed}}, \bibinfo {author} {\bibfnamefont {L.}~\bibnamefont {Motl}},
  \bibinfo {author} {\bibfnamefont {A.}~\bibnamefont {Nicolis}}, \ and\
  \bibinfo {author} {\bibfnamefont {C.}~\bibnamefont {Vafa}},\ }\href {\doibase
  10.1088/1126-6708/2007/06/060} {\bibfield  {journal} {\bibinfo  {journal}
  {JHEP}\ }\textbf {\bibinfo {volume} {06}},\ \bibinfo {pages} {060} (\bibinfo
  {year} {2007})},\ \Eprint {http://arxiv.org/abs/hep-th/0601001}
  {arXiv:hep-th/0601001 [hep-th]} \BibitemShut {NoStop}%
\bibitem [{\citenamefont {Cheung}\ and\ \citenamefont
  {Remmen}(2014)}]{Cheung:2014ega}%
  \BibitemOpen
  \bibfield  {author} {\bibinfo {author} {\bibfnamefont {C.}~\bibnamefont
  {Cheung}}\ and\ \bibinfo {author} {\bibfnamefont {G.~N.}\ \bibnamefont
  {Remmen}},\ }\href {\doibase 10.1007/JHEP12(2014)087} {\bibfield  {journal}
  {\bibinfo  {journal} {JHEP}\ }\textbf {\bibinfo {volume} {12}},\ \bibinfo
  {pages} {087} (\bibinfo {year} {2014})},\ \Eprint
  {http://arxiv.org/abs/1407.7865} {arXiv:1407.7865 [hep-th]} \BibitemShut
  {NoStop}%
\bibitem [{\citenamefont {Harlow}(2016)}]{Harlow:2015lma}%
  \BibitemOpen
  \bibfield  {author} {\bibinfo {author} {\bibfnamefont {D.}~\bibnamefont
  {Harlow}},\ }\href {\doibase 10.1007/JHEP01(2016)122} {\bibfield  {journal}
  {\bibinfo  {journal} {JHEP}\ }\textbf {\bibinfo {volume} {01}},\ \bibinfo
  {pages} {122} (\bibinfo {year} {2016})},\ \Eprint
  {http://arxiv.org/abs/1510.07911} {arXiv:1510.07911 [hep-th]} \BibitemShut
  {NoStop}%
\bibitem [{\citenamefont {Shiu}\ \emph {et~al.}(2016)\citenamefont {Shiu},
  \citenamefont {Soler},\ and\ \citenamefont {Cottrell}}]{Cottrell:2016bty}%
  \BibitemOpen
  \bibfield  {author} {\bibinfo {author} {\bibfnamefont {G.}~\bibnamefont
  {Shiu}}, \bibinfo {author} {\bibfnamefont {P.}~\bibnamefont {Soler}}, \ and\
  \bibinfo {author} {\bibfnamefont {W.}~\bibnamefont {Cottrell}},\ }\href@noop
  {} {\  (\bibinfo {year} {2016})},\ \Eprint {http://arxiv.org/abs/1611.06270}
  {arXiv:1611.06270 [hep-th]} \BibitemShut {NoStop}%
\bibitem [{\citenamefont {Hod}(2017)}]{Hod:2017uqc}%
  \BibitemOpen
  \bibfield  {author} {\bibinfo {author} {\bibfnamefont {S.}~\bibnamefont
  {Hod}},\ }\href {\doibase 10.1142/S0218271817420044} {\bibfield  {journal}
  {\bibinfo  {journal} {Int. J. Mod. Phys.}\ }\textbf {\bibinfo {volume}
  {D26}},\ \bibinfo {pages} {1742004} (\bibinfo {year} {2017})},\ \Eprint
  {http://arxiv.org/abs/1705.06287} {arXiv:1705.06287 [gr-qc]} \BibitemShut
  {NoStop}%
\bibitem [{\citenamefont {Fisher}\ and\ \citenamefont
  {Mogni}(2017)}]{Fisher:2017dbc}%
  \BibitemOpen
  \bibfield  {author} {\bibinfo {author} {\bibfnamefont {Z.}~\bibnamefont
  {Fisher}}\ and\ \bibinfo {author} {\bibfnamefont {C.~J.}\ \bibnamefont
  {Mogni}},\ }\href@noop {} {\  (\bibinfo {year} {2017})},\ \Eprint
  {http://arxiv.org/abs/1706.08257} {arXiv:1706.08257 [hep-th]} \BibitemShut
  {NoStop}%
\bibitem [{\citenamefont {Crisford}\ \emph {et~al.}(2018)\citenamefont
  {Crisford}, \citenamefont {Horowitz},\ and\ \citenamefont
  {Santos}}]{Crisford:2017gsb}%
  \BibitemOpen
  \bibfield  {author} {\bibinfo {author} {\bibfnamefont {T.}~\bibnamefont
  {Crisford}}, \bibinfo {author} {\bibfnamefont {G.~T.}\ \bibnamefont
  {Horowitz}}, \ and\ \bibinfo {author} {\bibfnamefont {J.~E.}\ \bibnamefont
  {Santos}},\ }\href {\doibase 10.1103/PhysRevD.97.066005} {\bibfield
  {journal} {\bibinfo  {journal} {Phys. Rev.}\ }\textbf {\bibinfo {volume}
  {D97}},\ \bibinfo {pages} {066005} (\bibinfo {year} {2018})},\ \Eprint
  {http://arxiv.org/abs/1709.07880} {arXiv:1709.07880 [hep-th]} \BibitemShut
  {NoStop}%
\bibitem [{\citenamefont {Cheung}\ \emph {et~al.}(2018)\citenamefont {Cheung},
  \citenamefont {Liu},\ and\ \citenamefont {Remmen}}]{Cheung:2018cwt}%
  \BibitemOpen
  \bibfield  {author} {\bibinfo {author} {\bibfnamefont {C.}~\bibnamefont
  {Cheung}}, \bibinfo {author} {\bibfnamefont {J.}~\bibnamefont {Liu}}, \ and\
  \bibinfo {author} {\bibfnamefont {G.~N.}\ \bibnamefont {Remmen}},\ }\href
  {\doibase 10.1007/JHEP10(2018)004} {\bibfield  {journal} {\bibinfo  {journal}
  {JHEP}\ }\textbf {\bibinfo {volume} {10}},\ \bibinfo {pages} {004} (\bibinfo
  {year} {2018})},\ \Eprint {http://arxiv.org/abs/1801.08546} {arXiv:1801.08546
  [hep-th]} \BibitemShut {NoStop}%
\bibitem [{\citenamefont {Hamada}\ \emph {et~al.}(2018)\citenamefont {Hamada},
  \citenamefont {Noumi},\ and\ \citenamefont {Shiu}}]{Hamada:2018dde}%
  \BibitemOpen
  \bibfield  {author} {\bibinfo {author} {\bibfnamefont {Y.}~\bibnamefont
  {Hamada}}, \bibinfo {author} {\bibfnamefont {T.}~\bibnamefont {Noumi}}, \
  and\ \bibinfo {author} {\bibfnamefont {G.}~\bibnamefont {Shiu}},\ }\href@noop
  {} {\  (\bibinfo {year} {2018})},\ \Eprint {http://arxiv.org/abs/1810.03637}
  {arXiv:1810.03637 [hep-th]} \BibitemShut {NoStop}%
\bibitem [{\citenamefont {Urbano}(2018)}]{Urbano:2018kax}%
  \BibitemOpen
  \bibfield  {author} {\bibinfo {author} {\bibfnamefont {A.}~\bibnamefont
  {Urbano}},\ }\href@noop {} {\  (\bibinfo {year} {2018})},\ \Eprint
  {http://arxiv.org/abs/1810.05621} {arXiv:1810.05621 [hep-th]} \BibitemShut
  {NoStop}%
\bibitem [{\citenamefont {Palti}(2017)}]{Palti:2017elp}%
  \BibitemOpen
  \bibfield  {author} {\bibinfo {author} {\bibfnamefont {E.}~\bibnamefont
  {Palti}},\ }\href {\doibase 10.1007/JHEP08(2017)034} {\bibfield  {journal}
  {\bibinfo  {journal} {JHEP}\ }\textbf {\bibinfo {volume} {08}},\ \bibinfo
  {pages} {034} (\bibinfo {year} {2017})},\ \Eprint
  {http://arxiv.org/abs/1705.04328} {arXiv:1705.04328 [hep-th]} \BibitemShut
  {NoStop}%
\bibitem [{\citenamefont {Lust}\ and\ \citenamefont
  {Palti}(2018)}]{Lust:2017wrl}%
  \BibitemOpen
  \bibfield  {author} {\bibinfo {author} {\bibfnamefont {D.}~\bibnamefont
  {Lust}}\ and\ \bibinfo {author} {\bibfnamefont {E.}~\bibnamefont {Palti}},\
  }\href {\doibase 10.1007/JHEP02(2018)040} {\bibfield  {journal} {\bibinfo
  {journal} {JHEP}\ }\textbf {\bibinfo {volume} {02}},\ \bibinfo {pages} {040}
  (\bibinfo {year} {2018})},\ \Eprint {http://arxiv.org/abs/1709.01790}
  {arXiv:1709.01790 [hep-th]} \BibitemShut {NoStop}%
\bibitem [{\citenamefont {Lee}\ \emph {et~al.}(2018)\citenamefont {Lee},
  \citenamefont {Lerche},\ and\ \citenamefont {Weigand}}]{Lee:2018urn}%
  \BibitemOpen
  \bibfield  {author} {\bibinfo {author} {\bibfnamefont {S.-J.}\ \bibnamefont
  {Lee}}, \bibinfo {author} {\bibfnamefont {W.}~\bibnamefont {Lerche}}, \ and\
  \bibinfo {author} {\bibfnamefont {T.}~\bibnamefont {Weigand}},\ }\href
  {\doibase 10.1007/JHEP10(2018)164} {\bibfield  {journal} {\bibinfo  {journal}
  {JHEP}\ }\textbf {\bibinfo {volume} {10}},\ \bibinfo {pages} {164} (\bibinfo
  {year} {2018})},\ \Eprint {http://arxiv.org/abs/1808.05958} {arXiv:1808.05958
  [hep-th]} \BibitemShut {NoStop}%
\bibitem [{\citenamefont {Palti}(2019)}]{Palti:2019pca}%
  \BibitemOpen
  \bibfield  {author} {\bibinfo {author} {\bibfnamefont {E.}~\bibnamefont
  {Palti}}\ }(\bibinfo {year} {2019})\ \Eprint
  {http://arxiv.org/abs/1903.06239} {arXiv:1903.06239 [hep-th]} \BibitemShut
  {NoStop}%
\bibitem [{\citenamefont {Gonzalo}\ and\ \citenamefont
  {Ibáñez}(2019)}]{Gonzalo:2019gjp}%
  \BibitemOpen
  \bibfield  {author} {\bibinfo {author} {\bibfnamefont {E.}~\bibnamefont
  {Gonzalo}}\ and\ \bibinfo {author} {\bibfnamefont {L.~E.}\ \bibnamefont
  {Ibáñez}},\ }\href@noop {} {\  (\bibinfo {year} {2019})},\ \Eprint
  {http://arxiv.org/abs/1903.08878} {arXiv:1903.08878 [hep-th]} \BibitemShut
  {NoStop}%
\bibitem [{\citenamefont {Yamazaki}(2019)}]{Yamazaki:2019ahj}%
  \BibitemOpen
  \bibfield  {author} {\bibinfo {author} {\bibfnamefont {M.}~\bibnamefont
  {Yamazaki}},\ }in\ \href@noop {} {\emph {\bibinfo {booktitle} {{54th
  Rencontres de Moriond on Electroweak Interactions and Unified Theories
  (Moriond EW 2019) La Thuile, Italy, March 16-23, 2019}}}}\ (\bibinfo {year}
  {2019})\ \Eprint {http://arxiv.org/abs/1904.05357} {arXiv:1904.05357
  [hep-ph]} \BibitemShut {NoStop}%
\bibitem [{\citenamefont {Misner}\ and\ \citenamefont
  {Wheeler}(1957)}]{Misner:1957mt}%
  \BibitemOpen
  \bibfield  {author} {\bibinfo {author} {\bibfnamefont {C.~W.}\ \bibnamefont
  {Misner}}\ and\ \bibinfo {author} {\bibfnamefont {J.~A.}\ \bibnamefont
  {Wheeler}},\ }\href {\doibase 10.1016/0003-4916(57)90049-0} {\bibfield
  {journal} {\bibinfo  {journal} {Annals Phys.}\ }\textbf {\bibinfo {volume}
  {2}},\ \bibinfo {pages} {525} (\bibinfo {year} {1957})}\BibitemShut {NoStop}%
\bibitem [{\citenamefont {Polchinski}(2004)}]{Polchinski:2003bq}%
  \BibitemOpen
  \bibfield  {author} {\bibinfo {author} {\bibfnamefont {J.}~\bibnamefont
  {Polchinski}},\ }\bibfield  {booktitle} {\emph {\bibinfo {booktitle}
  {{Proceedings, Dirac Centennial Symposium, Tallahassee, USA, December 6-7,
  2002}}},\ }\href {\doibase 10.1142/S0217751X0401866X} {\bibfield  {journal}
  {\bibinfo  {journal} {Int. J. Mod. Phys.}\ }\textbf {\bibinfo {volume}
  {A19S1}},\ \bibinfo {pages} {145} (\bibinfo {year} {2004})},\ \bibinfo {note}
  {[,145(2003)]},\ \Eprint {http://arxiv.org/abs/hep-th/0304042}
  {arXiv:hep-th/0304042 [hep-th]} \BibitemShut {NoStop}%
\bibitem [{\citenamefont {Banks}\ and\ \citenamefont
  {Seiberg}(2011)}]{Banks:2010zn}%
  \BibitemOpen
  \bibfield  {author} {\bibinfo {author} {\bibfnamefont {T.}~\bibnamefont
  {Banks}}\ and\ \bibinfo {author} {\bibfnamefont {N.}~\bibnamefont
  {Seiberg}},\ }\href {\doibase 10.1103/PhysRevD.83.084019} {\bibfield
  {journal} {\bibinfo  {journal} {Phys. Rev.}\ }\textbf {\bibinfo {volume}
  {D83}},\ \bibinfo {pages} {084019} (\bibinfo {year} {2011})},\ \Eprint
  {http://arxiv.org/abs/1011.5120} {arXiv:1011.5120 [hep-th]} \BibitemShut
  {NoStop}%
\bibitem [{\citenamefont {Arvanitaki}\ \emph {et~al.}(2010)\citenamefont
  {Arvanitaki}, \citenamefont {Dimopoulos}, \citenamefont {Dubovsky},
  \citenamefont {Kaloper},\ and\ \citenamefont
  {March-Russell}}]{Arvanitaki:2009fg}%
  \BibitemOpen
  \bibfield  {author} {\bibinfo {author} {\bibfnamefont {A.}~\bibnamefont
  {Arvanitaki}}, \bibinfo {author} {\bibfnamefont {S.}~\bibnamefont
  {Dimopoulos}}, \bibinfo {author} {\bibfnamefont {S.}~\bibnamefont
  {Dubovsky}}, \bibinfo {author} {\bibfnamefont {N.}~\bibnamefont {Kaloper}}, \
  and\ \bibinfo {author} {\bibfnamefont {J.}~\bibnamefont {March-Russell}},\
  }\href {\doibase 10.1103/PhysRevD.81.123530} {\bibfield  {journal} {\bibinfo
  {journal} {Phys. Rev.}\ }\textbf {\bibinfo {volume} {D81}},\ \bibinfo {pages}
  {123530} (\bibinfo {year} {2010})},\ \Eprint {http://arxiv.org/abs/0905.4720}
  {arXiv:0905.4720 [hep-th]} \BibitemShut {NoStop}%
\bibitem [{\citenamefont {Ratra}\ and\ \citenamefont
  {Peebles}(1988)}]{Ratra:1987rm}%
  \BibitemOpen
  \bibfield  {author} {\bibinfo {author} {\bibfnamefont {B.}~\bibnamefont
  {Ratra}}\ and\ \bibinfo {author} {\bibfnamefont {P.~J.~E.}\ \bibnamefont
  {Peebles}},\ }\href {\doibase 10.1103/PhysRevD.37.3406} {\bibfield  {journal}
  {\bibinfo  {journal} {Phys. Rev.}\ }\textbf {\bibinfo {volume} {D37}},\
  \bibinfo {pages} {3406} (\bibinfo {year} {1988})}\BibitemShut {NoStop}%
\bibitem [{\citenamefont {Wetterich}(1988)}]{Wetterich:1987fm}%
  \BibitemOpen
  \bibfield  {author} {\bibinfo {author} {\bibfnamefont {C.}~\bibnamefont
  {Wetterich}},\ }\href {\doibase 10.1016/0550-3213(88)90193-9} {\bibfield
  {journal} {\bibinfo  {journal} {Nucl. Phys.}\ }\textbf {\bibinfo {volume}
  {B302}},\ \bibinfo {pages} {668} (\bibinfo {year} {1988})},\ \Eprint
  {http://arxiv.org/abs/1711.03844} {arXiv:1711.03844 [hep-th]} \BibitemShut
  {NoStop}%
\bibitem [{\citenamefont {Zlatev}\ \emph {et~al.}(1999)\citenamefont {Zlatev},
  \citenamefont {Wang},\ and\ \citenamefont {Steinhardt}}]{Zlatev:1998tr}%
  \BibitemOpen
  \bibfield  {author} {\bibinfo {author} {\bibfnamefont {I.}~\bibnamefont
  {Zlatev}}, \bibinfo {author} {\bibfnamefont {L.-M.}\ \bibnamefont {Wang}}, \
  and\ \bibinfo {author} {\bibfnamefont {P.~J.}\ \bibnamefont {Steinhardt}},\
  }\href {\doibase 10.1103/PhysRevLett.82.896} {\bibfield  {journal} {\bibinfo
  {journal} {Phys. Rev. Lett.}\ }\textbf {\bibinfo {volume} {82}},\ \bibinfo
  {pages} {896} (\bibinfo {year} {1999})},\ \Eprint
  {http://arxiv.org/abs/astro-ph/9807002} {arXiv:astro-ph/9807002 [astro-ph]}
  \BibitemShut {NoStop}%
\bibitem [{\citenamefont {Fischbach}\ and\ \citenamefont
  {Talmadge}(1999)}]{Fischbach:1999bc}%
  \BibitemOpen
  \bibfield  {author} {\bibinfo {author} {\bibfnamefont {E.}~\bibnamefont
  {Fischbach}}\ and\ \bibinfo {author} {\bibfnamefont {C.~L.}\ \bibnamefont
  {Talmadge}},\ }\href@noop {} {\emph {\bibinfo {title} {{The search for
  nonNewtonian gravity}}}}\ (\bibinfo {year} {1999})\BibitemShut {NoStop}%
\bibitem [{\citenamefont {Bertotti}\ \emph {et~al.}(2003)\citenamefont
  {Bertotti}, \citenamefont {Iess},\ and\ \citenamefont
  {Tortora}}]{Bertotti:2003rm}%
  \BibitemOpen
  \bibfield  {author} {\bibinfo {author} {\bibfnamefont {B.}~\bibnamefont
  {Bertotti}}, \bibinfo {author} {\bibfnamefont {L.}~\bibnamefont {Iess}}, \
  and\ \bibinfo {author} {\bibfnamefont {P.}~\bibnamefont {Tortora}},\ }\href
  {\doibase 10.1038/nature01997} {\bibfield  {journal} {\bibinfo  {journal}
  {Nature}\ }\textbf {\bibinfo {volume} {425}},\ \bibinfo {pages} {374}
  (\bibinfo {year} {2003})}\BibitemShut {NoStop}%
\bibitem [{\citenamefont {Esposito-Farese}(2004)}]{EspositoFarese:2004cc}%
  \BibitemOpen
  \bibfield  {author} {\bibinfo {author} {\bibfnamefont {G.}~\bibnamefont
  {Esposito-Farese}},\ }\bibfield  {booktitle} {\emph {\bibinfo {booktitle}
  {{Phi in the sky: The quest for cosmological scalar fields. Proceedings,
  Workshop, Porto, Portugal, July 8-10, 2004}}},\ }\href {\doibase
  10.1063/1.1835173} {\bibfield  {journal} {\bibinfo  {journal} {AIP Conf.
  Proc.}\ }\textbf {\bibinfo {volume} {736}},\ \bibinfo {pages} {35} (\bibinfo
  {year} {2004})},\ \Eprint {http://arxiv.org/abs/gr-qc/0409081}
  {arXiv:gr-qc/0409081 [gr-qc]} \BibitemShut {NoStop}%
\bibitem [{\citenamefont {Will}(2006)}]{Will:2005va}%
  \BibitemOpen
  \bibfield  {author} {\bibinfo {author} {\bibfnamefont {C.~M.}\ \bibnamefont
  {Will}},\ }\href {\doibase 10.12942/lrr-2006-3} {\bibfield  {journal}
  {\bibinfo  {journal} {Living Rev. Rel.}\ }\textbf {\bibinfo {volume} {9}},\
  \bibinfo {pages} {3} (\bibinfo {year} {2006})},\ \Eprint
  {http://arxiv.org/abs/gr-qc/0510072} {arXiv:gr-qc/0510072 [gr-qc]}
  \BibitemShut {NoStop}%
\bibitem [{\citenamefont {Obied}\ \emph {et~al.}(2018)\citenamefont {Obied},
  \citenamefont {Ooguri}, \citenamefont {Spodyneiko},\ and\ \citenamefont
  {Vafa}}]{Obied:2018sgi}%
  \BibitemOpen
  \bibfield  {author} {\bibinfo {author} {\bibfnamefont {G.}~\bibnamefont
  {Obied}}, \bibinfo {author} {\bibfnamefont {H.}~\bibnamefont {Ooguri}},
  \bibinfo {author} {\bibfnamefont {L.}~\bibnamefont {Spodyneiko}}, \ and\
  \bibinfo {author} {\bibfnamefont {C.}~\bibnamefont {Vafa}},\ }\href@noop {}
  {\  (\bibinfo {year} {2018})},\ \Eprint {http://arxiv.org/abs/1806.08362}
  {arXiv:1806.08362 [hep-th]} \BibitemShut {NoStop}%
\bibitem [{\citenamefont {Denef}\ \emph {et~al.}(2018)\citenamefont {Denef},
  \citenamefont {Hebecker},\ and\ \citenamefont {Wrase}}]{Denef:2018etk}%
  \BibitemOpen
  \bibfield  {author} {\bibinfo {author} {\bibfnamefont {F.}~\bibnamefont
  {Denef}}, \bibinfo {author} {\bibfnamefont {A.}~\bibnamefont {Hebecker}}, \
  and\ \bibinfo {author} {\bibfnamefont {T.}~\bibnamefont {Wrase}},\ }\href
  {\doibase 10.1103/PhysRevD.98.086004} {\bibfield  {journal} {\bibinfo
  {journal} {Phys. Rev.}\ }\textbf {\bibinfo {volume} {D98}},\ \bibinfo {pages}
  {086004} (\bibinfo {year} {2018})},\ \Eprint
  {http://arxiv.org/abs/1807.06581} {arXiv:1807.06581 [hep-th]} \BibitemShut
  {NoStop}%
\bibitem [{\citenamefont {Conlon}(2018)}]{Conlon:2018eyr}%
  \BibitemOpen
  \bibfield  {author} {\bibinfo {author} {\bibfnamefont {J.~P.}\ \bibnamefont
  {Conlon}},\ }\href {\doibase 10.1142/S0217751X18501786} {\bibfield  {journal}
  {\bibinfo  {journal} {Int. J. Mod. Phys.}\ }\textbf {\bibinfo {volume}
  {A33}},\ \bibinfo {pages} {1850178} (\bibinfo {year} {2018})},\ \Eprint
  {http://arxiv.org/abs/1808.05040} {arXiv:1808.05040 [hep-th]} \BibitemShut
  {NoStop}%
\bibitem [{\citenamefont {Murayama}\ \emph {et~al.}(2018)\citenamefont
  {Murayama}, \citenamefont {Yamazaki},\ and\ \citenamefont
  {Yanagida}}]{Murayama:2018lie}%
  \BibitemOpen
  \bibfield  {author} {\bibinfo {author} {\bibfnamefont {H.}~\bibnamefont
  {Murayama}}, \bibinfo {author} {\bibfnamefont {M.}~\bibnamefont {Yamazaki}},
  \ and\ \bibinfo {author} {\bibfnamefont {T.~T.}\ \bibnamefont {Yanagida}},\
  }\href {\doibase 10.1007/JHEP12(2018)032} {\bibfield  {journal} {\bibinfo
  {journal} {JHEP}\ }\textbf {\bibinfo {volume} {12}},\ \bibinfo {pages} {032}
  (\bibinfo {year} {2018})},\ \Eprint {http://arxiv.org/abs/1809.00478}
  {arXiv:1809.00478 [hep-th]} \BibitemShut {NoStop}%
\bibitem [{\citenamefont {Choi}\ \emph {et~al.}(2018)\citenamefont {Choi},
  \citenamefont {Chway},\ and\ \citenamefont {Shin}}]{Choi:2018rze}%
  \BibitemOpen
  \bibfield  {author} {\bibinfo {author} {\bibfnamefont {K.}~\bibnamefont
  {Choi}}, \bibinfo {author} {\bibfnamefont {D.}~\bibnamefont {Chway}}, \ and\
  \bibinfo {author} {\bibfnamefont {C.~S.}\ \bibnamefont {Shin}},\ }\href
  {\doibase 10.1007/JHEP11(2018)142} {\bibfield  {journal} {\bibinfo  {journal}
  {JHEP}\ }\textbf {\bibinfo {volume} {11}},\ \bibinfo {pages} {142} (\bibinfo
  {year} {2018})},\ \Eprint {http://arxiv.org/abs/1809.01475} {arXiv:1809.01475
  [hep-th]} \BibitemShut {NoStop}%
\bibitem [{\citenamefont {Hamaguchi}\ \emph {et~al.}(2018)\citenamefont
  {Hamaguchi}, \citenamefont {Ibe},\ and\ \citenamefont
  {Moroi}}]{Hamaguchi:2018vtv}%
  \BibitemOpen
  \bibfield  {author} {\bibinfo {author} {\bibfnamefont {K.}~\bibnamefont
  {Hamaguchi}}, \bibinfo {author} {\bibfnamefont {M.}~\bibnamefont {Ibe}}, \
  and\ \bibinfo {author} {\bibfnamefont {T.}~\bibnamefont {Moroi}},\ }\href
  {\doibase 10.1007/JHEP12(2018)023} {\bibfield  {journal} {\bibinfo  {journal}
  {JHEP}\ }\textbf {\bibinfo {volume} {12}},\ \bibinfo {pages} {023} (\bibinfo
  {year} {2018})},\ \Eprint {http://arxiv.org/abs/1810.02095} {arXiv:1810.02095
  [hep-th]} \BibitemShut {NoStop}%
\bibitem [{\citenamefont {Dvali}\ and\ \citenamefont
  {Gomez}(2019)}]{Dvali:2018fqu}%
  \BibitemOpen
  \bibfield  {author} {\bibinfo {author} {\bibfnamefont {G.}~\bibnamefont
  {Dvali}}\ and\ \bibinfo {author} {\bibfnamefont {C.}~\bibnamefont {Gomez}},\
  }\href {\doibase 10.1002/prop.201800092} {\bibfield  {journal} {\bibinfo
  {journal} {Fortsch. Phys.}\ }\textbf {\bibinfo {volume} {67}},\ \bibinfo
  {pages} {1800092} (\bibinfo {year} {2019})},\ \Eprint
  {http://arxiv.org/abs/1806.10877} {arXiv:1806.10877 [hep-th]} \BibitemShut
  {NoStop}%
\bibitem [{\citenamefont {Andriot}(2018)}]{Andriot:2018wzk}%
  \BibitemOpen
  \bibfield  {author} {\bibinfo {author} {\bibfnamefont {D.}~\bibnamefont
  {Andriot}},\ }\href {\doibase 10.1016/j.physletb.2018.09.022} {\bibfield
  {journal} {\bibinfo  {journal} {Phys. Lett.}\ }\textbf {\bibinfo {volume}
  {B785}},\ \bibinfo {pages} {570} (\bibinfo {year} {2018})},\ \Eprint
  {http://arxiv.org/abs/1806.10999} {arXiv:1806.10999 [hep-th]} \BibitemShut
  {NoStop}%
\bibitem [{\citenamefont {Garg}\ and\ \citenamefont
  {Krishnan}(2018)}]{Garg:2018reu}%
  \BibitemOpen
  \bibfield  {author} {\bibinfo {author} {\bibfnamefont {S.~K.}\ \bibnamefont
  {Garg}}\ and\ \bibinfo {author} {\bibfnamefont {C.}~\bibnamefont
  {Krishnan}},\ }\href@noop {} {\  (\bibinfo {year} {2018})},\ \Eprint
  {http://arxiv.org/abs/1807.05193} {arXiv:1807.05193 [hep-th]} \BibitemShut
  {NoStop}%
\bibitem [{\citenamefont {Ooguri}\ \emph {et~al.}(2019)\citenamefont {Ooguri},
  \citenamefont {Palti}, \citenamefont {Shiu},\ and\ \citenamefont
  {Vafa}}]{Ooguri:2018wrx}%
  \BibitemOpen
  \bibfield  {author} {\bibinfo {author} {\bibfnamefont {H.}~\bibnamefont
  {Ooguri}}, \bibinfo {author} {\bibfnamefont {E.}~\bibnamefont {Palti}},
  \bibinfo {author} {\bibfnamefont {G.}~\bibnamefont {Shiu}}, \ and\ \bibinfo
  {author} {\bibfnamefont {C.}~\bibnamefont {Vafa}},\ }\href {\doibase
  10.1016/j.physletb.2018.11.018} {\bibfield  {journal} {\bibinfo  {journal}
  {Phys. Lett.}\ }\textbf {\bibinfo {volume} {B788}},\ \bibinfo {pages} {180}
  (\bibinfo {year} {2019})},\ \Eprint {http://arxiv.org/abs/1810.05506}
  {arXiv:1810.05506 [hep-th]} \BibitemShut {NoStop}%
\bibitem [{\citenamefont {Garg}\ \emph {et~al.}(2019)\citenamefont {Garg},
  \citenamefont {Krishnan},\ and\ \citenamefont {Zaid~Zaz}}]{Garg:2018zdg}%
  \BibitemOpen
  \bibfield  {author} {\bibinfo {author} {\bibfnamefont {S.~K.}\ \bibnamefont
  {Garg}}, \bibinfo {author} {\bibfnamefont {C.}~\bibnamefont {Krishnan}}, \
  and\ \bibinfo {author} {\bibfnamefont {M.}~\bibnamefont {Zaid~Zaz}},\ }\href
  {\doibase 10.1007/JHEP03(2019)029} {\bibfield  {journal} {\bibinfo  {journal}
  {JHEP}\ }\textbf {\bibinfo {volume} {03}},\ \bibinfo {pages} {029} (\bibinfo
  {year} {2019})},\ \Eprint {http://arxiv.org/abs/1810.09406} {arXiv:1810.09406
  [hep-th]} \BibitemShut {NoStop}%
\bibitem [{\citenamefont {Andriot}\ and\ \citenamefont
  {Roupec}(2019)}]{Andriot:2018mav}%
  \BibitemOpen
  \bibfield  {author} {\bibinfo {author} {\bibfnamefont {D.}~\bibnamefont
  {Andriot}}\ and\ \bibinfo {author} {\bibfnamefont {C.}~\bibnamefont
  {Roupec}},\ }\href {\doibase 10.1002/prop.201800105} {\bibfield  {journal}
  {\bibinfo  {journal} {Fortsch. Phys.}\ }\textbf {\bibinfo {volume} {67}},\
  \bibinfo {pages} {1800105} (\bibinfo {year} {2019})},\ \Eprint
  {http://arxiv.org/abs/1811.08889} {arXiv:1811.08889 [hep-th]} \BibitemShut
  {NoStop}%
\bibitem [{\citenamefont {Dine}\ and\ \citenamefont
  {Seiberg}(1985)}]{Dine:1985he}%
  \BibitemOpen
  \bibfield  {author} {\bibinfo {author} {\bibfnamefont {M.}~\bibnamefont
  {Dine}}\ and\ \bibinfo {author} {\bibfnamefont {N.}~\bibnamefont {Seiberg}},\
  }\href {\doibase 10.1016/0370-2693(85)90927-X} {\bibfield  {journal}
  {\bibinfo  {journal} {Phys. Lett.}\ }\textbf {\bibinfo {volume} {162B}},\
  \bibinfo {pages} {299} (\bibinfo {year} {1985})}\BibitemShut {NoStop}%
\bibitem [{\citenamefont {Maldacena}\ and\ \citenamefont
  {Nunez}(2001)}]{Maldacena:2000mw}%
  \BibitemOpen
  \bibfield  {author} {\bibinfo {author} {\bibfnamefont {J.~M.}\ \bibnamefont
  {Maldacena}}\ and\ \bibinfo {author} {\bibfnamefont {C.}~\bibnamefont
  {Nunez}},\ }\bibfield  {booktitle} {\emph {\bibinfo {booktitle}
  {{Superstrings. Proceedings, International Conference, Strings 2000, Ann
  Arbor, USA, July 10-15, 2000}}},\ }\href {\doibase 10.1142/S0217751X01003935,
  10.1142/S0217751X01003937} {\bibfield  {journal} {\bibinfo  {journal} {Int.
  J. Mod. Phys.}\ }\textbf {\bibinfo {volume} {A16}},\ \bibinfo {pages} {822}
  (\bibinfo {year} {2001})},\ \bibinfo {note} {[,182(2000)]},\ \Eprint
  {http://arxiv.org/abs/hep-th/0007018} {arXiv:hep-th/0007018 [hep-th]}
  \BibitemShut {NoStop}%
\bibitem [{\citenamefont {Steinhardt}\ and\ \citenamefont
  {Wesley}(2009)}]{Steinhardt:2008nk}%
  \BibitemOpen
  \bibfield  {author} {\bibinfo {author} {\bibfnamefont {P.~J.}\ \bibnamefont
  {Steinhardt}}\ and\ \bibinfo {author} {\bibfnamefont {D.}~\bibnamefont
  {Wesley}},\ }\href {\doibase 10.1103/PhysRevD.79.104026} {\bibfield
  {journal} {\bibinfo  {journal} {Phys. Rev.}\ }\textbf {\bibinfo {volume}
  {D79}},\ \bibinfo {pages} {104026} (\bibinfo {year} {2009})},\ \Eprint
  {http://arxiv.org/abs/0811.1614} {arXiv:0811.1614 [hep-th]} \BibitemShut
  {NoStop}%
\bibitem [{\citenamefont {McOrist}\ and\ \citenamefont
  {Sethi}(2012)}]{McOrist:2012yc}%
  \BibitemOpen
  \bibfield  {author} {\bibinfo {author} {\bibfnamefont {J.}~\bibnamefont
  {McOrist}}\ and\ \bibinfo {author} {\bibfnamefont {S.}~\bibnamefont
  {Sethi}},\ }\href {\doibase 10.1007/JHEP12(2012)122} {\bibfield  {journal}
  {\bibinfo  {journal} {JHEP}\ }\textbf {\bibinfo {volume} {12}},\ \bibinfo
  {pages} {122} (\bibinfo {year} {2012})},\ \Eprint
  {http://arxiv.org/abs/1208.0261} {arXiv:1208.0261 [hep-th]} \BibitemShut
  {NoStop}%
\bibitem [{\citenamefont {Sethi}(2018)}]{Sethi:2017phn}%
  \BibitemOpen
  \bibfield  {author} {\bibinfo {author} {\bibfnamefont {S.}~\bibnamefont
  {Sethi}},\ }\href {\doibase 10.1007/JHEP10(2018)022} {\bibfield  {journal}
  {\bibinfo  {journal} {JHEP}\ }\textbf {\bibinfo {volume} {10}},\ \bibinfo
  {pages} {022} (\bibinfo {year} {2018})},\ \Eprint
  {http://arxiv.org/abs/1709.03554} {arXiv:1709.03554 [hep-th]} \BibitemShut
  {NoStop}%
\bibitem [{\citenamefont {Danielsson}\ and\ \citenamefont
  {Van~Riet}(2018)}]{Danielsson:2018ztv}%
  \BibitemOpen
  \bibfield  {author} {\bibinfo {author} {\bibfnamefont {U.~H.}\ \bibnamefont
  {Danielsson}}\ and\ \bibinfo {author} {\bibfnamefont {T.}~\bibnamefont
  {Van~Riet}},\ }\href {\doibase 10.1142/S0218271818300070} {\bibfield
  {journal} {\bibinfo  {journal} {Int. J. Mod. Phys.}\ }\textbf {\bibinfo
  {volume} {D27}},\ \bibinfo {pages} {1830007} (\bibinfo {year} {2018})},\
  \Eprint {http://arxiv.org/abs/1804.01120} {arXiv:1804.01120 [hep-th]}
  \BibitemShut {NoStop}%
\bibitem [{\citenamefont {Klaewer}\ and\ \citenamefont
  {Palti}(2017)}]{Klaewer:2016kiy}%
  \BibitemOpen
  \bibfield  {author} {\bibinfo {author} {\bibfnamefont {D.}~\bibnamefont
  {Klaewer}}\ and\ \bibinfo {author} {\bibfnamefont {E.}~\bibnamefont
  {Palti}},\ }\href {\doibase 10.1007/JHEP01(2017)088} {\bibfield  {journal}
  {\bibinfo  {journal} {JHEP}\ }\textbf {\bibinfo {volume} {01}},\ \bibinfo
  {pages} {088} (\bibinfo {year} {2017})},\ \Eprint
  {http://arxiv.org/abs/1610.00010} {arXiv:1610.00010 [hep-th]} \BibitemShut
  {NoStop}%
\bibitem [{\citenamefont {Brahma}\ and\ \citenamefont
  {Hossain}(2019)}]{Brahma:2019mdd}%
  \BibitemOpen
  \bibfield  {author} {\bibinfo {author} {\bibfnamefont {S.}~\bibnamefont
  {Brahma}}\ and\ \bibinfo {author} {\bibfnamefont {M.~W.}\ \bibnamefont
  {Hossain}},\ }\href@noop {} {\  (\bibinfo {year} {2019})},\ \Eprint
  {http://arxiv.org/abs/1904.05810} {arXiv:1904.05810 [hep-th]} \BibitemShut
  {NoStop}%
\bibitem [{\citenamefont {Degrassi}\ \emph {et~al.}(2012)\citenamefont
  {Degrassi}, \citenamefont {Di~Vita}, \citenamefont {Elias-Miro},
  \citenamefont {Espinosa}, \citenamefont {Giudice}, \citenamefont {Isidori},\
  and\ \citenamefont {Strumia}}]{Degrassi:2012ry}%
  \BibitemOpen
  \bibfield  {author} {\bibinfo {author} {\bibfnamefont {G.}~\bibnamefont
  {Degrassi}}, \bibinfo {author} {\bibfnamefont {S.}~\bibnamefont {Di~Vita}},
  \bibinfo {author} {\bibfnamefont {J.}~\bibnamefont {Elias-Miro}}, \bibinfo
  {author} {\bibfnamefont {J.~R.}\ \bibnamefont {Espinosa}}, \bibinfo {author}
  {\bibfnamefont {G.~F.}\ \bibnamefont {Giudice}}, \bibinfo {author}
  {\bibfnamefont {G.}~\bibnamefont {Isidori}}, \ and\ \bibinfo {author}
  {\bibfnamefont {A.}~\bibnamefont {Strumia}},\ }\href {\doibase
  10.1007/JHEP08(2012)098} {\bibfield  {journal} {\bibinfo  {journal} {JHEP}\
  }\textbf {\bibinfo {volume} {08}},\ \bibinfo {pages} {098} (\bibinfo {year}
  {2012})},\ \Eprint {http://arxiv.org/abs/1205.6497} {arXiv:1205.6497
  [hep-ph]} \BibitemShut {NoStop}%
\bibitem [{\citenamefont {Buttazzo}\ \emph {et~al.}(2013)\citenamefont
  {Buttazzo}, \citenamefont {Degrassi}, \citenamefont {Giardino}, \citenamefont
  {Giudice}, \citenamefont {Sala}, \citenamefont {Salvio},\ and\ \citenamefont
  {Strumia}}]{Buttazzo:2013uya}%
  \BibitemOpen
  \bibfield  {author} {\bibinfo {author} {\bibfnamefont {D.}~\bibnamefont
  {Buttazzo}}, \bibinfo {author} {\bibfnamefont {G.}~\bibnamefont {Degrassi}},
  \bibinfo {author} {\bibfnamefont {P.~P.}\ \bibnamefont {Giardino}}, \bibinfo
  {author} {\bibfnamefont {G.~F.}\ \bibnamefont {Giudice}}, \bibinfo {author}
  {\bibfnamefont {F.}~\bibnamefont {Sala}}, \bibinfo {author} {\bibfnamefont
  {A.}~\bibnamefont {Salvio}}, \ and\ \bibinfo {author} {\bibfnamefont
  {A.}~\bibnamefont {Strumia}},\ }\href {\doibase 10.1007/JHEP12(2013)089}
  {\bibfield  {journal} {\bibinfo  {journal} {JHEP}\ }\textbf {\bibinfo
  {volume} {12}},\ \bibinfo {pages} {089} (\bibinfo {year} {2013})},\ \Eprint
  {http://arxiv.org/abs/1307.3536} {arXiv:1307.3536 [hep-ph]} \BibitemShut
  {NoStop}%
\bibitem [{\citenamefont {Tanabashi}\ \emph {et~al.}(2018)\citenamefont
  {Tanabashi} \emph {et~al.}}]{Tanabashi:2018oca}%
  \BibitemOpen
  \bibfield  {author} {\bibinfo {author} {\bibfnamefont {M.}~\bibnamefont
  {Tanabashi}} \emph {et~al.} (\bibinfo {collaboration} {Particle Data
  Group}),\ }\href {\doibase 10.1103/PhysRevD.98.030001} {\bibfield  {journal}
  {\bibinfo  {journal} {Phys. Rev.}\ }\textbf {\bibinfo {volume} {D98}},\
  \bibinfo {pages} {030001} (\bibinfo {year} {2018})}\BibitemShut {NoStop}%
\bibitem [{\citenamefont {Nagata}\ and\ \citenamefont
  {Shirai}(2014)}]{Nagata:2013sba}%
  \BibitemOpen
  \bibfield  {author} {\bibinfo {author} {\bibfnamefont {N.}~\bibnamefont
  {Nagata}}\ and\ \bibinfo {author} {\bibfnamefont {S.}~\bibnamefont
  {Shirai}},\ }\href {\doibase 10.1007/JHEP03(2014)049} {\bibfield  {journal}
  {\bibinfo  {journal} {JHEP}\ }\textbf {\bibinfo {volume} {03}},\ \bibinfo
  {pages} {049} (\bibinfo {year} {2014})},\ \Eprint
  {http://arxiv.org/abs/1312.7854} {arXiv:1312.7854 [hep-ph]} \BibitemShut
  {NoStop}%
\bibitem [{\citenamefont {Vainshtein}(1972)}]{Vainshtein:1972sx}%
  \BibitemOpen
  \bibfield  {author} {\bibinfo {author} {\bibfnamefont {A.~I.}\ \bibnamefont
  {Vainshtein}},\ }\href {\doibase 10.1016/0370-2693(72)90147-5} {\bibfield
  {journal} {\bibinfo  {journal} {Phys. Lett.}\ }\textbf {\bibinfo {volume}
  {39B}},\ \bibinfo {pages} {393} (\bibinfo {year} {1972})}\BibitemShut
  {NoStop}%
\bibitem [{\citenamefont {Khoury}\ and\ \citenamefont
  {Weltman}(2004)}]{Khoury:2003rn}%
  \BibitemOpen
  \bibfield  {author} {\bibinfo {author} {\bibfnamefont {J.}~\bibnamefont
  {Khoury}}\ and\ \bibinfo {author} {\bibfnamefont {A.}~\bibnamefont
  {Weltman}},\ }\href {\doibase 10.1103/PhysRevD.69.044026} {\bibfield
  {journal} {\bibinfo  {journal} {Phys. Rev.}\ }\textbf {\bibinfo {volume}
  {D69}},\ \bibinfo {pages} {044026} (\bibinfo {year} {2004})},\ \Eprint
  {http://arxiv.org/abs/astro-ph/0309411} {arXiv:astro-ph/0309411 [astro-ph]}
  \BibitemShut {NoStop}%
\bibitem [{\citenamefont {Hinterbichler}\ \emph {et~al.}(2011)\citenamefont
  {Hinterbichler}, \citenamefont {Khoury},\ and\ \citenamefont
  {Nastase}}]{Hinterbichler:2010wu}%
  \BibitemOpen
  \bibfield  {author} {\bibinfo {author} {\bibfnamefont {K.}~\bibnamefont
  {Hinterbichler}}, \bibinfo {author} {\bibfnamefont {J.}~\bibnamefont
  {Khoury}}, \ and\ \bibinfo {author} {\bibfnamefont {H.}~\bibnamefont
  {Nastase}},\ }\href {\doibase 10.1007/JHEP06(2011)072,
  10.1007/JHEP03(2011)061} {\bibfield  {journal} {\bibinfo  {journal} {JHEP}\
  }\textbf {\bibinfo {volume} {03}},\ \bibinfo {pages} {061} (\bibinfo {year}
  {2011})},\ \bibinfo {note} {[Erratum: JHEP06,072(2011)]},\ \Eprint
  {http://arxiv.org/abs/1012.4462} {arXiv:1012.4462 [hep-th]} \BibitemShut
  {NoStop}%
\bibitem [{\citenamefont {Adams}\ \emph {et~al.}(2006)\citenamefont {Adams},
  \citenamefont {Arkani-Hamed}, \citenamefont {Dubovsky}, \citenamefont
  {Nicolis},\ and\ \citenamefont {Rattazzi}}]{Adams:2006sv}%
  \BibitemOpen
  \bibfield  {author} {\bibinfo {author} {\bibfnamefont {A.}~\bibnamefont
  {Adams}}, \bibinfo {author} {\bibfnamefont {N.}~\bibnamefont {Arkani-Hamed}},
  \bibinfo {author} {\bibfnamefont {S.}~\bibnamefont {Dubovsky}}, \bibinfo
  {author} {\bibfnamefont {A.}~\bibnamefont {Nicolis}}, \ and\ \bibinfo
  {author} {\bibfnamefont {R.}~\bibnamefont {Rattazzi}},\ }\href {\doibase
  10.1088/1126-6708/2006/10/014} {\bibfield  {journal} {\bibinfo  {journal}
  {JHEP}\ }\textbf {\bibinfo {volume} {10}},\ \bibinfo {pages} {014} (\bibinfo
  {year} {2006})},\ \Eprint {http://arxiv.org/abs/hep-th/0602178}
  {arXiv:hep-th/0602178 [hep-th]} \BibitemShut {NoStop}%
\end{thebibliography}%
\end{document}